%% file: autosam.tex
\documentclass[twocolumn]{autart}    

\usepackage{graphicx}          

\usepackage{cite}
\usepackage{amsmath,amssymb,amsfonts}
\usepackage{algorithmic}
\usepackage{graphicx}
\usepackage{cuted}
\usepackage{flushend}
\usepackage{mathtools}
\usepackage{array}
\usepackage{subcaption}
\mathtoolsset{showonlyrefs}
\usepackage{multirow}
\usepackage{multicol}
\usepackage{xcolor}
\usepackage{braket}
\usepackage{floatrow}
\usepackage{kantlipsum}
\usepackage{enumitem}
\usepackage{hyphenat}
\usepackage{enumitem}
\usepackage{wrapfig}

\newtheorem{theorem}{Theorem}[section]
\newtheorem{definition}{Definition}[section]
\newtheorem{assumption}{Assumption}[section]
\newtheorem{corollary}{Corollary}[theorem]
\newtheorem{lemma}[theorem]{Lemma}
\newtheorem{remark}{Remark}
\newtheorem{problem}{Problem}

\definecolor{ajgCol}{rgb}{0,0,0}
\newcommand{\ajg}[1]{{\color{ajgCol}#1}}

\definecolor{scaCol}{rgb}{0,0,0}

\newcommand{\PP}{\mathcal{P}}
\newcommand{\CC}{\mathcal{C}}
\newcommand{\WW}{\mathcal{W}}
\newcommand{\QQ}{\mathcal{Q}}
\newcommand{\HH}{\mathcal{H}}
\newcommand{\GG}{\mathcal{G}}

\usepackage[round]{natbib}        
\allowdisplaybreaks

\begin{document}

\begin{frontmatter}

\title{Switching Multiplicative Watermark Design \\ Against Covert Attacks\thanksref{footnoteinfo}} 

\thanks[footnoteinfo]{This work has been partially supported by the Research Council of Norway through the project AIMWind (grant ID 312486), by the Swedish Research Council under the grant 2018-04396, and by the Swedish Foundation for Strategic Research. The material in this paper was partially presented at the 60th IEEE Conference on Decision and Control, Austin, Texas, 2021. Corresponding author Sribalaji. C. Anand. 
}

\author[Paestum]{Alexander J. Gallo\thanksref{footnoteinfo2}}\ead{alexanderjulian.gallo@polimi.it},    
\author[Rome]{Sribalaji C. Anand\thanksref{footnoteinfo2}}\ead{srca@kth.se},               
\author[Baiae]{Andre M. H. Teixeira}\ead{andre.teixeira@it.uu.se},  
\author[Pompeii]{Riccardo M. G. Ferrari}\ead{r.ferrari@tudelft.nl}

\address[Paestum]{Department of Electronics, Information and Bioengineering, Politecnico di Milano, Milano, Italy.}  
\address[Rome]{School of Electrical Engineering and Computer Science and Digital Futures, KTH Royal Institute of Technology, Sweden}             
\address[Baiae]{Department of Information Technology, Uppsala University, PO Box 337, SE-75105, Uppsala, Sweden.}        

\address[Pompeii]{Delft Center for Systems and Control, Mechanical Engineering, TU Delft, Delft, Netherlands.}

\thanks[footnoteinfo2]{These authors contributed equally.}
          

\begin{abstract}                          
\textit{Active techniques} have been introduced to give better detectability performance for cyber-attack diagnosis in cyber-physical systems (CPS). In this paper, switching multiplicative watermarking is considered, 
whereby we propose an optimal design strategy to define switching filter parameters. 
Optimality is evaluated exploiting the so-called output-to-output gain of the closed loop system, including some supposed attack dynamics. 
A worst-case scenario of a matched covert attack is assumed, presuming that an attacker with full knowledge of the closed-loop system injects a stealthy attack of bounded energy. 
Our algorithm, given watermark filter parameters at some time instant, provides optimal next-step parameters.
Analysis of the algorithm is given, demonstrating its features, and demonstrating that through initialization of certain parameters outside of the algorithm, the parameters of the multiplicative watermarking can be randomized.
Simulation shows how, by adopting our method for parameter design, the attacker's impact on performance diminishes.
\end{abstract}

\end{frontmatter}

\section{Introduction}
\input{Introduction}

\section{Problem description}\label{sec:PF}

\input{PF.tex}
%
\section{Optimal design of filters}\label{sec:main}
\input{design.tex}
\section{Numerical example}\label{sec:NE}
\input{NE_updated}

\section{Conclusion and future works }\label{sec:Con}
An optimal design technique for the design of the parameters of switching multiplicative watermarking filters is presented. 
The problem is formalized by supposing the closed-loop system is subject to a covert attack with matching parameters. 
We propose an optimal control problem based on a formulation of the attack energy constrained output-to-output gain. 
We show through a numerical example that this design improves detectability by increasing the energy of the residual output before and after a switching event. 
Future works includes developing algorithms for optimal design and optimal switching times ensuring that mWM does not destabilize the closed-loop system under switching with mismatched parameters, and studying non-linear systems. 
\bibliography{autosam_abbrv}

\end{document}

%% file: Introduction.tex
The widespread integration of communication networks and smart devices in modern control systems has increased the vulnerability of industrial systems to online cyber-attacks, e.g., Industroyer, Blackenergy, etc \citep{osti_1505628}.
To counter this, methods have been developed to improve security by achieving attack detection, mitigation, and monitoring, among others \citep{sandberg2022secure}. This paper focuses on active attack diagnosis to mitigate stealthy attacks. 
%

Active diagnosis techniques rely on the inclusion of additional moduli to control systems
to alter the behavior of the system compared to information known by the attacker. 
For instance, the concept of additive watermarking was introduced in \cite{mo2015physical}, where noise signals of known mean and variance are added at the plant and compensated for it at the controller. 
This compensation, however, is not exact, causing some performance degradation. Thus, trade-offs between performance and detectability  are necessary \citep{zhu2023detection}.

In encrypted control \citep{darup2021encrypted}, the sensor data is encrypted, sent to the controller, and then operated on directly. Encrypted input signals are sent back to the plant for decryption. Although encryption is widespread in IT security, in control systems it presents some concerns, such as the introduction of time delays \citep{stabile2024verifiable}, while it may present inherent weaknesses \citep{alisic2023model}.

In moving target defense \citep{griffioen2020moving}, the plant is augmented with fictitious dynamics, known to the controller. The plant output is transmitted to the controller along with the fictitious states over a network under attack. 
The additional measurements then aide in the detection of attacks. 
This comes at the cost of higher communication bandwidth needs, which increases rapidly with the dimension of the augmented systems.

Other recently proposed works include two-way coding \citep{fang2019two}, a weak encryuption technique, and dynamic masking \citep{abdalmoaty2023privacy}, which enhances privacy as well as security, have been shown to be effective against zero-dynamics attacks.
Furthermore, filtering techniques for attack detection are proposed by \cite{murguia2020security,hashemi2022codesign,escudero2023safety}, while not focusing on stealthy attacks.

Multiplicative watermarking (mWM) has been proposed by the authors as a diagnosis technique \citep{ferrari2020switching}. mWM consists of a pair of filters on each communication channel between the plant and its controller; the scheme is affine to weak encryption, whereby ``encoding'' and ``decoding'' are done by changing signals' dynamic characteristics through inverse pairs of filters. This enables original signals to be recovered exactly, and thus does not lead to performance degradation.

One of the critical features of multiplicative watermarking is that to detect stealthy attacks, the mWM filter parameters must be switched over time. In this paper, an algorithm to optimally design the mWM parameters after a switching event is presented, enhancing detection performance, without changing the switching time.

\textcolor{black}{
To formalize the filter design problem, we suppose the defender is interested in optimal performance against adversaries injecting covert attacks with matched system parameters \citep{smith2015covert}, including the mWM parameters prior to the switch. This scenario represents a worst case where malicious agents can take full control of the system while remaining undetected.
Thus, the attack strategy is explicitly included within the formulation of the closed-loop system, and the mWM filters are chosen by solving an optimization problem minimizing the attack-energy-constrained output-to-output gain (AEC-OOG) \citep{anand2023risk}, a variation of the output-to-output gain proposed in  \cite{teixeira2015strategic}.
}
The main contributions of this paper are:
\begin{enumerate}
\item The problem of optimally designing the switching mWM filters is formulated as an optimization problem, with the AEC-OOG is taken as the objective;
\item The worst-case scenario of a covert attack with exact knowledge of plant and mWM filter parameters is embedded within the design problem;
\item The feasibility of the optimization problem is shown to be dependent only on stability conditions; 
\item A solution scheme is proposed to promote randomization of the mWM filter parameters such that an eavesdropping adversary cannot remain stealthy.
\end{enumerate} 

This builds on the results of \cite{ferrari2020switching}, where the focus was on the design of the switching protocols, rather than the parameters themselves.
Compared to previous work \citep{gallo2021design}, this paper introduces an optimization problem which is always feasible (thanks to the use of AEC-OOG in the objective), while also considering a more sophisticated class of covert attacks, where the presence of watermark is known to the adversary. 
Moreover, this paper poses a different objective than \citep{zhang2023hybrid}; indeed, while \citep{zhang2023hybrid} provided a design strategy to ensure certain privacy properties, in this paper we address the problem of optimal parameter design following a switching event.

The rest of the paper is organized as follows. 
After formulating the problem in Section~\ref{sec:PF}, we propose our design algorithm in Section~\ref{sec:main}, and analyze its properties. It is then evaluated through a numerical example in Section~\ref{sec:NE}, and concluding remarks are given Section~\ref{sec:Con}.

%% file: PF.tex
\begin{figure}
    \centering
    \includegraphics[width=6.5cm]{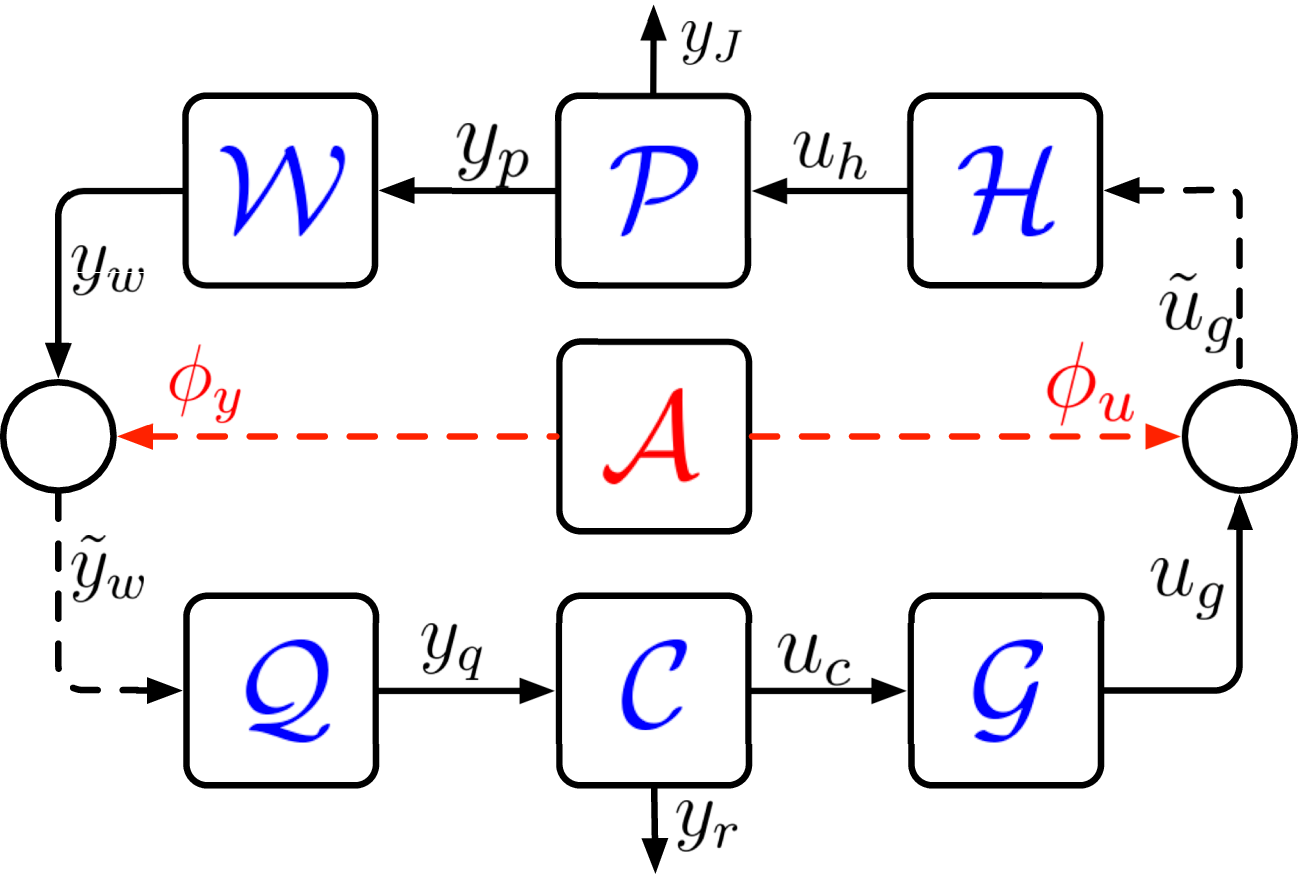}
    \caption{Block diagram of the closed-loop CPS including the plant $\PP$, controller $\CC$ and watermarking filters $\{\WW,\QQ,\GG,\HH\}$. The information transmitted between $\PP$ and $\CC$ is altered by the adversary $\mathcal A$. The dashed lines represent the network affected by the adversary.}
    \label{fig:sys}
\end{figure}

We consider the Cyber-Physical System (CPS) in Fig.~\ref{fig:sys}. This includes plant $\PP$, controller and anomaly detector $\CC$, mWM filters $\WW,\QQ,\GG,\HH$, and the malicious agent $\mathcal{A}$. The mWM filters are defined pairwise, namely $\{\QQ,\WW\}$ and $\{\GG,\HH\}$ are referred to as, respectively, the output and input \textit{mWM filter pairs}. 

\subsection{Plant and controller}
Consider an LTI discrete-time (DT) plant modeled by: 
	\begin{equation}\label{eq:sys}
	    \PP: \left\{ \begin{aligned}
	        x_p[k+1] &= A_p x_p[k] + B_p u_h[k]\\
	        y_p[k] &= C_p x_p[k]\\
            y_J[k] &= C_J x_p[k] + D_J u_h[k]
	    \end{aligned}\right.
	\end{equation}
	where $x_p \in \mathbb R^n$ is the plant's state, $u_h \in \mathbb R^m$ its input, $y_p \in \mathbb R^p$ its measured output, and all the system's matrices are of the appropriate dimension. Furthermore, suppose a (possibly unmeasured) \textit{performance output} $y_J \in \mathbb{R}^{p_J}$ is defined, such that the performance of the system, evaluated over the interval $[k-N+1,k]$, for some $N \in \mathbb{N}$ \citep{zhou1996robust}, is given by:
	\begin{align}
	    J(x_p,u_h) &= 
        \|y_J\|^2_{\ell_2,[k-N+1,k]}.
	\end{align}
    \begin{assumption}
    The tuples $(A_p,B_p)$ and $(C_p,A_p)$ are respectively, controllable and observable pairs.
		$\hfill\triangleleft$
	\end{assumption}
 \begin{assumption}\label{ass:stable}
     The plant $\mathcal{P}$ is stable and $x_p[0]=0$. $\hfill\triangleleft$
 \end{assumption}

Assumption~\ref{ass:stable}, necessary for the OOG to be meaningful \citep{teixeira2015secure}, does not reduce generality, as stability can be ensured by a local (non-networked) controller \citep{hu2007stability,lin2023secondary}, whilst $x_p[0] = 0$ can be considered because of linearity.


The plant is regulated by an observer-based dynamic controller $\CC$, described by:
\begin{equation}\label{eq:cntrl}
    \mathcal C: \left\{
	\begin{aligned}
		\hat{x}_p[k+1] &= A_p \hat{x}_p[k] + B_p u_c[k] + Ly_r[k]\\
		u_c[k] &= K\hat{x}_p[k]\\
		\hat{y}_p[k] &= C_p \hat{x}_p[k]\\
		y_r[k] &= y_q[k] - \hat{y}_p[k]
	\end{aligned}\right.
\end{equation}
where $\hat{x}_p \in \mathbb{R}^n, \hat{y}_p \in \mathbb{R}^p$ are the state and measurement estimates, $u_c \in \mathbb{R}^m$ the control input. The matrices $K$ and $L$ are the controller and observer gains respectively. Finally, the term $y_r$ in \eqref{eq:cntrl} is the residual output, used to detect the presence of an attack: given a threshold $\epsilon_r$, an attack is detected if the inequality $\|y_r\|_{\ell_2,[0,N_r]}^2 \leq \epsilon_r$ is falsified for any $N_r \in \mathbb{N}_+$. Note that in \eqref{eq:sys}-\eqref{eq:cntrl} $y_q$ and $u_h$, the outputs of $\QQ$ and $\HH$ (to be defined), are used as the input to the controller and the plant, respectively. 

\subsection{Multiplicative watermarking filters}
Consider mWM filters defined as follows
\begin{equation}\label{eq:sys:WM}
		\Sigma: \begin{cases}
			x_\sigma[k+1] = A_\sigma(\theta_\sigma[k]) x_\sigma [k] + B_\sigma(\theta_\sigma[k]) \nu_\sigma [k]\\
			\gamma_\sigma [k] = C_\sigma(\theta_\sigma[k]) x_\sigma [k] + D_\sigma(\theta_\sigma[k]) \nu_\sigma [k]
		\end{cases},
	\end{equation}
	with $\Sigma \in \{\GG,\HH,\WW,\QQ\}$, $\sigma \in \{g,h,w,q\}$, \ajg{where $g,h,w,q$ refer to variables pertaining to $\mathcal G, \mathcal H, \mathcal W, \mathcal Q$, respectively}\footnote{In the sequel whenever referring to the parameters of any one of the mWM filters, the subscript $\sigma$ is used. Conversely, if referring to all parameters, $\theta$ is used.}, $x_\sigma \in \mathbb{R}^{n_\sigma}$ the state of $\Sigma$, $\nu_\sigma \in \mathbb{R}^{m_\sigma}$ its input, $\gamma_\sigma \in \mathbb{R}^{p_\sigma}$ the output, and $\theta_{\sigma}[k]$ is a vector of parameters.

\begin{definition}[mWM filter parameters]
    The parameter $\theta_\sigma[k]$ is taken to be the concatenation of the vectorized form of all matrices $A_\sigma(\cdot), B_\sigma(\cdot), C_\sigma(\cdot), D_\sigma(\cdot)$.
    $\hfill\triangleleft$
\end{definition}

The parameter $\theta_{\sigma}$ is defined to be piecewise constant:
$$\theta_\sigma[k] = \bar{\theta}_\sigma[k_i], \forall k \in \{k_i, k_i+1, \dots, k_{i+1}-1\}$$
where $k_i, i = 0,1, \dots \in \mathbb{N}_+,$ are switching instants. In the following, with some abuse of notation, the time dependencies are dropped, with $\theta_\sigma$ and $\theta_\sigma^+$ used to define the parameters before and after a switching instant, 
i.e., $\theta_\sigma = \theta_\sigma[k_i]$, $\theta_\sigma^+ = \theta_\sigma[k_{i+1}]$.


Furthermore, all filters are taken to be square systems, i.e., $m_\sigma = p_\sigma, \forall \sigma \in \{g,h,w,q\}$, and define $\nu_g \triangleq u_c, \nu_h \triangleq \tilde{u}_g, \nu_w \triangleq y_p, \nu_q \triangleq \tilde{y}_w, \gamma_g \triangleq u_g, \gamma_h \triangleq u_h, \gamma_w \triangleq y_w, \gamma_q \triangleq y_q.$
    Here, a \textit{tilde} is used to highlight that $\tilde u_g, \tilde y_w$ are received through the insecure communication network and as such may be affected by attacks. 
%
%
\begin{remark}
    The objective of this paper is to \textit{optimally} design the successive parameters of the mWM filters $\theta_\sigma^+$, given their value $\theta_\sigma$. 
    It remains out of the scope of the paper to address other aspects of the switching mechanisms, such as determining the switching time, or defining the jump functions for the states. 
    Interested readers are referred to \citep{ferrari2020switching}.
    $\hfill \triangleleft$
\end{remark}

\begin{definition}[Watermarking pair]\label{def:WM}
Two systems $(\mathcal W,\mathcal Q)$ \eqref{eq:sys:WM}, are a \textit{watermarking pair} if:
\begin{enumerate}[label=\alph*.]
    \item \label{def:WM:inv} $\mathcal{W}$ and $\mathcal{Q}$ are stable and invertible, i.e., exists a positive definite matrix $Z_\sigma \succ 0, \sigma \in \{w,q\}$ such that 
    \begin{equation}\label{eq:WM:stab}
        A_\sigma^\top Z_\sigma A_\sigma - Z_\sigma \prec 0;
    \end{equation}
    \item \label{def:WM:eqParam} if $\theta_w[k] = \theta_q[k]$, $y_q[k] = y_p[k]$, i.e., 
    \begin{equation}\label{eq:WM:def}
	   \QQ \triangleq \WW^{-1}\,. \qquad \triangleleft
    \end{equation}
\end{enumerate}
\end{definition}
\begin{remark}\label{rem:stabSw}
    If $Z_\sigma$ in \eqref{eq:WM:stab} is the same for all $\theta_\sigma[k], k \in \mathbb N,\sigma\in\{g,h,w,q\}$, the mWM filters, on their own, are stable under arbitrary switching, as they all share a common Lyapunov function.
    $\hfill \triangleleft$
\end{remark}

\begin{definition}[{\cite[Lemma 3.15]{zhou1996robust}}]\label{def:inv:ss}
		Define the DT transfer function resulting from the system defined by the tuple $(A,B,C,D)$ as $			G(z) = \left[\begin{array}{c|c}
				A &B\\
				\hline
				C &D
			\end{array}
			\right],$
		and suppose that $D^{-1}$ exists. Then
		\begin{equation}\label{eq:sys:inv}
			G^{-1}(z) = \left[\begin{array}{c|c}
				A - BD^{-1}C    &BD^{-1}\\
				\hline
				-D^{-1}C        &D^{-1}
			\end{array}
			\right]
		\end{equation}
		is the inverse transfer function of $G(z)$.
		$\hfill\triangleleft$
	\end{definition}
\begin{assumption}\label{ass:sync}
    The mWM parameters are matched, i.e., $\theta_w[k] = \theta_q[k]$ and $\theta_g[k] = \theta_h[k]\; \forall\;k \in \mathbb{N}$.
    $\hfill \triangleleft$
\end{assumption}

\subsection{Attack model}\label{ch:probFor:atk}
Consider the malicious agent $\mathcal A$ located in the CPS as in Fig.~\ref{fig:sys}, capable of tampering with data transmitted between $\PP$ and $\CC$. 
Without loss of generality, the injected attacks are modeled as additive signals:
 \begin{equation}\label{eq:atk}
     \tilde{u}_g[k] \triangleq u_g[k] + \varphi_u[k],\;\;\; \tilde{y}_w[k] \triangleq y_w[k] + \varphi_y[k],
 \end{equation}
where $\varphi_u[k]$ and $\varphi_y[k]$ are actuator and sensor attack signals designed by the adversary $\mathcal A$. To properly define our design algorithm in Section~\ref{sec:main}, an explicit strategy for the attack signals $\varphi_u$ and $\varphi_y$ must be defined by the defender. 
In this paper, we focus on covert attacks \citep{smith2015covert}, which remain undetected for passive diagnosis scheme.


The covert attack strategy, under Assumption~\ref{ass:param} and~\ref{ass:atkEng}, is as follows: the malicious agent $\mathcal A$ chooses $\varphi_u[k] \in \ell_{2e}$ freely, while $\varphi_y[k]$ satisfies:
\begin{equation}\label{eq:atk:cov}
	\mathcal A:
\left\{
\begin{aligned}
    x_a[k+1] &= A_a(\theta^a) x_a[k] + B_a(\theta^a) \varphi_u[k]\\
		y_a[k] &= C_a(\theta^a) x_a[k] + D_a(\theta^a) \varphi_u[k]\\
		\varphi_y[k] &= - y_a[k]
\end{aligned}
\right.
\end{equation}
where $x_a \triangleq [x_{h,a}^\top\;x_{p,a}^\top\;x_{w,a}^\top]^\top$ is the attacker's state, and its dynamics are the same as the cascade of $\HH, \PP, \WW$, parametrized\footnote{\ajg{Here, and throughout the paper, a super- or subscript $a$ is used to indicate that a variable pertains to $\mathcal A$.}} by $\theta^a_\sigma$.

\begin{assumption}\label{ass:param}
For all $k \in [k_{i+1},k_{i+2}], i \in \mathbb N_+$, the attacker parameters $\theta_\sigma^a[k]=\theta_\sigma[k_i]$, $\sigma \in \{h,w\}. \hfill\triangleleft$
\end{assumption}

\begin{assumption}\label{ass:atkEng}
    The attack energy is bounded and finite, i.e.,: $\Vert \varphi_u\Vert_{\ell_2}^2 \leq \epsilon_a$, with $\epsilon_a$ known to $\mathcal C$. $\hfill \triangleleft$
\end{assumption}

\begin{remark}
    Assumption~\ref{ass:atkEng} is introduced as it allows for guarantees that the algorithm proposed in Section~\ref{sec:main} always returns a feasible solution (see Theorem~\ref{thm:well:posed}).
    In general, 
    while it may be that the adversary has limited energy \citep{djouadi2015finite}, it is a strong assumption that the bound $\epsilon_a$ is known to the defender.
    Nonetheless, the attack energy bound $\epsilon_a$ may be seen as a design variable that, together with the chosen attack model \eqref{eq:atk:cov}, facilitates the definition of a systematic design of mWM filters by the defender.
    %
    %
    Further remarks regarding the consequences of Assumption~\ref{ass:atkEng} not holding are postponed to Remark~\ref{rem:atkEng2}, following the formal definition of the attack-energy-constrained output-to-output gain in Definition~\ref{def:o2o}. 
    $\hfill \triangleleft$
\end{remark}

\subsection{Problem formulation}\label{sec:PF:probFor}
The objective of this paper is to propose a design strategy capable of optimally designing the mWM filter parameters $\theta^+$, supposing a covert attack is present within the CPS. To formulate a metric to be used to define optimality, the closed-loop CPS dynamics must be defined. Under the attack strategy \eqref{eq:atk:cov}, the closed-loop system with the attack $\varphi_u$ as input and the performance and detection output as system outputs can be rewritten as
\begin{equation}\label{eq:S_cl}
		{\mathcal{S}}:\left\{
\begin{aligned}
{x}[k+1] &= A x[k] + B\varphi_u[k]\\
y_J[k] &= \bar{C}_J x[k] + \bar{D}_J \varphi_u[k]\\
y_r[k] &= \bar{C}_r x[k]
\end{aligned} \right.
	\end{equation}
where $x = \begin{bmatrix}x_p^\top, &x_h^\top, &x_g^\top, &x_c^\top, &x_q^\top, &x_w^\top, &x_a^\top\end{bmatrix}^\top$ is the closed-loop system state, while $y_r$ and $y_J$ remain the residual and performance outputs. All signals in \eqref{eq:S_cl} are also a function of the parameters $\theta^+$, but this dependence is dropped for clarity.
The definition of the matrices in \eqref{eq:S_cl} follow from \eqref{eq:sys}-\eqref{eq:sys:WM} and \eqref{eq:atk:cov}.

The defender aims to quantify (and later minimize) the maximum performance loss caused by a stealthy and bounded-energy adversary on \eqref{eq:S_cl}. 
This is done by exploiting the attack-energy-constrained output-to-output gain 
(AEC-OOG) \citep{anand2023risk}. 
\begin{definition}[AEC-OOG]\label{def:o2o}
	The AEC-OOG
    of $\mathcal S$ in \eqref{eq:S_cl} is the value of the following optimization problem:
		\begin{equation}\label{eq:o2o}
			\begin{aligned}
				\sup_{\varphi_u\in\ell_{2e}} &\quad \|y_J\|_{\ell_2}^2 \\
				\text{s.t.}& \quad  \|y_r\|_{\ell_2}^2 \leq \epsilon_r,\; \|\varphi_u\|_{\ell_2}^2 \leq \epsilon_a,\;x[0] = 0.
			\end{aligned}
		\end{equation}
	where $\epsilon_a$ is the energy bound of the attack signal, $\epsilon_r$ is the detection threshold, and the value of \eqref{eq:o2o} denotes the performance loss caused by a stealthy adversary.	$\hfill\triangleleft$
	\end{definition}

\begin{problem}\label{problem_main}
Given $\theta_{\sigma}$ at some switching time $k_i, i \in \mathbb{N}_+$, find the optimal set of mWM filter parameters after a switching event $\theta_{\sigma}^+$, such that the AEC-OOG of the system $\mathcal{S}$ in \eqref{eq:S_cl} is minimized. $\hfill \triangleleft$
\end{problem}

\begin{remark}
    Because of its dependence on the AEC-OOG, the solution of Problem~\ref{problem_main} at time $k_i$ relies explicitly on the attack parameters $\theta^a[k_i]$. Given the malicious agent's strategy outlined in Section~\ref{ch:probFor:atk}, and Assumption~\ref{ass:param}, $\theta_\sigma^a[k_i] = \theta_\sigma^+[k_{i-1}]$ is known to $\CC$, without any additional knowledge required. 
    $\hfill \triangleleft$
\end{remark}

\begin{remark}\label{rem:atkEng2}
    We are now ready to formally treat the violation of Assumption~\ref{ass:atkEng}.
    To do this, let us first remark on some properties of the AEC-OOG, which follow from using finite bounds $\epsilon_r$ and $\epsilon_a$.
    Firstly, as demonstrated in Theorem~\ref{thm:well:posed}, the metric is always bounded, making it well suited for a design algorithm.
    Furthermore, it is explicitly related to both the $H_\infty$ metric and the original OOG proposed in \cite{teixeira2015strategic}, for increasing values of $\epsilon_r$ and $\epsilon_a$, respectively \citep[Prop.1]{anand2023risk}.
    Finally, we can comment on the constraint on the attack energy.
    Consider the value of \eqref{eq:o2o} under increasing values of $\epsilon_a$, as well the OOG as defined in \cite{teixeira2015strategic}.
    If the OOG is finite, there is some value $\bar\epsilon_a$ such that the AEC-OOG is the same as the OOG for all $\epsilon_a \geq \bar\epsilon_a$.
    If there are exploitable zero dynamcis, and the OOG is unbounded, $\|y_J\|_{\ell_2}^2$ grows unbounded as $\epsilon_a \rightarrow \infty$. Thus, while $\theta_\sigma^+$, the solution to Problem~\ref{problem_main}, is only optimal for covert attacks satisfying $\|\varphi_u\|_{\ell_2}^2 \leq \epsilon_a$, it ensures that the effect of $\varphi_u$ on $y_J$ is in some sense minimal if the attack energy constraint is violated.
    $\hfill\triangleleft$
\end{remark}



%% file: design.tex

\subsection{Design problem}
As summarized in Problem~\ref{problem_main}, the objective of the parameter design is to \textit{minimize the maximum performance loss caused by the adversary.} 
This can be translated, exploiting \eqref{eq:o2o}, to the following optimization problem
\begin{align}\label{eq:o2o:problem}
    \inf_{\theta^+} \mathcal{L}(\theta^+,\theta^a)
\end{align}
\begin{equation}\label{p1}
\mathcal{L}(\theta^+,\theta^a) =
 \left\{    \begin{aligned}
    \sup_{\varphi_u \in\ell_{2e}} & \; \|y_J\|_{\ell_2}^2 \\
		\text{s.t.} &\; \|y_r\|_{\ell_2}^2  \leq \epsilon_r,\; \|\varphi_u\|_{\ell_2}^2 \leq \epsilon_a,\\
                    &\; x[0] = 0, \eqref{eq:WM:stab}, \eqref{eq:WM:def},\eqref{eq:atk:cov},\eqref{eq:S_cl} 
\end{aligned}\right.
\end{equation}

In \eqref{eq:o2o:problem}, $\mathcal{L}(\theta^+,\theta^a)$ represents the value of the maximum performance loss caused by the adversary for any given pair of filters $(\theta^+,\theta^a)$.
The optimization problem \eqref{eq:o2o:problem} is an infinite optimization problem in signal space. Using \cite[Lem 3.1, Lem 3.2]{anand2023risk}, \eqref{eq:o2o:problem} is converted to an equivalent, finite-dimensional, non-convex optimization problem in Lemma \ref{lem:sig_2_mat}.
\begin{lemma}\label{lem:sig_2_mat}
The infinite-dimensional optimization problem \eqref{eq:o2o:problem} is equivalent to the following finite-dimensional, non-convex optimization problem
\begin{equation}\label{o1}
\begin{aligned}
\inf_{P,\gamma,\gamma_a,\theta^+,Z_\sigma} & \;\epsilon_r\gamma +\epsilon_a\gamma_a \\
\mathrm{s.t.} \;~ & R+\begin{bmatrix}
    \bar{C}_J^\top \bar{C}_J-\gamma \bar{C}_r^\top \bar{C}_r & \bar{C}_J^\top \bar{D}_J\\
    \bar{D}_J^\top \bar{C}_J & \bar{D}_J^\top \bar{D}_J^\top -\gamma_aI_{m}
\end{bmatrix} \preceq 0\\
& \eqref{eq:WM:stab}, \eqref{eq:WM:def}, \gamma \geq 0, \gamma_a \geq 0, P \succeq 0, Z_\sigma \succ 0,
\end{aligned}
\end{equation} 
where 
$R \triangleq \begin{bmatrix}
    A^\top PA-P & A^\top PB\\B^\top PA & B^\top PB
\end{bmatrix}$.
$\hfill \square$  
\end{lemma}
Finding a solution to \eqref{o1} solves Problem~\ref{problem_main}, as solving for $\theta^+$ achieves the minimal worst-case impact of a covert attack satisfying Assumption~\ref{ass:param}. Although \eqref{o1} is convex in $P,\gamma$ and $\gamma_a$, it contains non-convex terms in $A$.
As such, it cannot be easily solved via standard convex solvers \citep{lofberg2004yalmip}. 

\subsection{Well-posedness of the impact metric \eqref{eq:o2o:problem}}
Differently to our previous results~\citep{gallo2021design}, using the AEC-OOG ensures that the optimization problem used for the design of the mWM parameters is always feasible, as summarized in the following.


\begin{theorem}\label{thm:well:posed}
Let the parameters $\theta^{+}$ be chosen such that \eqref{eq:WM:stab}, \eqref{eq:WM:def}, and Assumption \ref{ass:stable} hold. 
Then, the value of the metric $\mathcal{L}$ in \eqref{p1} is bounded if the closed-loop matrix $A$ in \eqref{eq:S_cl} is Schur stable. 
$\hfill \square$
\end{theorem}

\begin{pf} 
Let $\Sigma_J \triangleq (A,B,\bar{C}_J,\bar{D}_J)$ be the closed loop system from the attack input $\varphi_u$ to the performance output $y_J$. The objective is to show that the value of \eqref{p1} is bounded given Assumption \ref{ass:stable}, and for any given value of $\theta^+$ that satisfies \eqref{eq:WM:stab} and \eqref{eq:WM:def}. To this end, start by considering the optimization problem \eqref{p1} without the constraint $||y_r||_{\ell_2}^2 \leq \epsilon_r$. 
The value of the resulting optimization problem is the $H_{\infty}$ gain of the system $\Sigma_J$, which is bounded, so long as $\Sigma_J$ is stable. Thus, \eqref{p1} is bounded, as the optimal value of any maximization problem cannot increase under additional constraints.
$\hfill\blacksquare$
\end{pf}
The condition of $A$ being stable is required only at any given time $k \in \mathbb N$, and not under switching. The problem of ensuring $A$ is stable under switching is addressed in \cite[Thm. 3]{ferrari2020switching}.

\subsection{Filter parameter update algorithm}\label{ch:design:algo}
As mentioned previously, the optimization problem \eqref{o1} is non-convex and cannot be solved exactly. 
One approach to solve \eqref{o1} is to reformulate the problem with Bi-linear Matrix inequalities (BMI) and use some existing approaches in the literature to solve them (e.g., \cite{gallo2021design,dehnert2021less,dinh2011combining}, etc.), which however come with drawbacks. In light of this, here an exhaustive search algorithm, defined in Algorithm~\ref{algo2}, is adopted, to show the main advantage of the proposed design problem \eqref{o1}.
%
%

The exhaustive search algorithm we proposed can be sketched out as follows. Let the values of all matrices be chose \textit{a priori}, apart from $A_\sigma$, such that they satisfy \eqref{eq:WM:def}.
Thus, the objective is to find optimal values of $A_\sigma$ minimizing \eqref{o1}. Furthermore, to ensure tractability, let us restrict the matrices $A_\sigma$ to be diagonal. To guarantee stability of the watermark generating matrices, it is sufficient to constrain the diagonal elements to lie in $(-1,1)$.
Discretizing this set into a grid of $n_s$ elements, $\mathcal A_h$ and $\mathcal A_q$ can be obtained, with cardinality $n_s^{n_h}$ and $n_s^{n_q}$, respectively.
Thus, the exhaustive search algorithm searches for optimal matrices $A_h, A_q$, under the constraint \eqref{eq:WM:def}.
The complete algorithm is summarized in Algorithm~\ref{algo2}, where the final step provides an ordering, in case multiple parameters obtain the same optimum.

\subsection{\ajg{Randomizing the solution}
}
\label{ch:design:nonRep}

Until now, the design of the algorithm has been purely deterministic: given the parameters $\theta$, \eqref{o1} uniquely determines the parameters $\theta^+$.
This provides optimal results, but it makes the architecture vulnerable to attacks\footnote{The attacker in question is different to that defined in Section~\ref{ch:probFor:atk}, where the attack strategy was considered as a \textit{design choice} for the formulation of the optimization problem.} capable of identifying the mWM filter parameters, as the attacker can compute future values of $\theta$ by solving Algorithm~\ref{algo2}.
We therefore propose a method to counteract this vulnerability. 
%
Specifically, by initializing matrices $D_q$ and $D_h$ randomly in the first step of the algorithm, it can be shown that the resulting parameters $\theta^+$ are also random.

\begin{algorithm}
\caption{Filter parameters selection algorithm}\vspace{-5pt}
\noindent\rule{8cm}{0.1pt}
\textbf{Initialization}: $K,L,\theta, \theta^a, \gamma^* = \infty$ \\
\textbf{Result:} $\theta^{+,*}$\vspace{-5pt}\\
\noindent\rule{8cm}{0.1pt}
\begin{enumerate}[label=\textbf{\arabic*:},leftmargin=*]
\item Pick random matrices $D_h$ and $D_q$.
\item[] \textbf{While} $((\mathcal{A}_h \neq \emptyset) || (\mathcal{A}_q \neq \emptyset))$, \textbf{do:}
    \item Draw a matrix $A_h$ from $\mathcal{A}_h$ and delete it from $\mathcal{A}_h$. 
    \item If the inverse of $A_h: A_g$ obtained from $\eqref{eq:WM:def}$ is unstable go to step $\textbf{2}$.
    \item Draw a matrix $A_q$ from $\mathcal{A}_q$ and delete it from $\mathcal{A}_q$. 
    \item If the inverse of $A_q: A_w$ obtained from $\eqref{eq:WM:def}$ is unstable go to step $\textbf{4}$.
    \item Derive the inverse filters using \eqref{eq:WM:def}.
    \item Using the values of the watermarking filters, and $\theta^a$, solve the convex optimization problem \eqref{o1}. Let us denote the value of \eqref{o1} as $\gamma_t$.
    \item If $\gamma_t < \gamma^*$, store the values of watermarking parameters, else go back to step (1).
    \item[] \textbf{end While}\vspace{-15pt}\\
\end{enumerate}
\noindent\rule{8cm}{0.1pt}
\label{algo2}
\end{algorithm}

\begin{theorem}\label{thm:nonRepeat}
    Let us $D_q[k_i],D_h[k_i]$ be the matrices defined in Step~1 of Algorithm~\ref{algo2} at switching times $k_i, i = 0,1,\dots \in \mathbb{N}_+$. It is sufficient to select $D_q[k_i] \neq D_q[k_j]$ and $D_h[k_i] \neq D_h[k_j]$ to ensure that $\theta_\sigma^+(k_i) \neq \theta_\sigma^+(k_j)$, $\forall i,j = 0,1,\dots \in \mathcal N_+, i \neq j$. 
    $\hfill\square$
\end{theorem}
\begin{pf}
    The proof follows directly from the fact that, for any two state space realizations $(A_1,B_1,C_1,D_1)$ and $(A_2,B_2,C_2,D_2)$ with compatible dimensions, it is sufficient for $D_1 \neq D_2$ for the resulting transfer functions $G_1(z) \neq G_2(z)$ \cite[Th. 4.1]{chen1984linear}.
    As a consequence, so long as $D_h[k_i] \neq D_h[k_j]$ and $D_q[k_i] \neq D_q[k_j]$, there are no mWM parameters such that the resulting closed loop transfer functions are the same.
    $\hfill\blacksquare$
\end{pf}


\begin{corollary}
    Let $D_h[k_i], D_q[k_i]$ be realizations of random variables. Then, the filter parameters $\theta^+[k_i]$ are also randomized.
    $\hfill\square$
\end{corollary}

To ensure that the parameters $\theta^+$ remain synchronous, it is necessary for the randomized values of $D_h, D_q$ be the same on both plant and control side.
The problem of selecting variables that are synchronized and (pseudo)random is a common issue in the secure control literature, and different solutions have been found, such as (\cite{zhang2022sensor}, \cite{zhang2023hybrid}).

%% file: NE_updated.tex
\subsection{Plant description}
Consider a power generating system \citep[Sec.4]{park2019stealthy} 
modeled by the dynamics:
\begin{align}
\label{power_AB} \begin{bmatrix}
\dot{\eta}_1\\ \dot{\eta}_2 \\ \dot{\eta}_3
\end{bmatrix} &= 
\begin{bmatrix}
\frac{-1}{T_{lm}} & \frac{K_{lm}}{T_{lm}} & \frac{-2K_{lm}}{T_{lm}}\\
0 & \frac{-2}{T_h} & \frac{6}{T_h}\\
\frac{-1}{T_g R} & 0 & \frac{-1}{T_g}
\end{bmatrix}
\underbrace{\begin{bmatrix}
{\eta}_1\\ {\eta}_2 \\ {\eta}_3
\end{bmatrix}}_{\eta}
+ \begin{bmatrix}
0\\ 0 \\ \frac{1}{T_g}
\end{bmatrix}
{u}\\
\label{power_C} y_p &= \underbrace{ \begin{bmatrix}
1 & 0 & 0 
\end{bmatrix}}_{C_p}\eta,\;\;
y_J = \underbrace{
\begin{bmatrix}
0 & 1 & 0
\end{bmatrix}}_{C_J}\eta.
\end{align}
Here, $\eta \triangleq [df; dp + 2 dx; dx]$, where $df$ is the frequency deviation in \mbox{Hz}, $dp$ is the change in the generator output per unit (\mbox{p.u.}), and $dx$ is the change in the valve position \mbox{p.u.}. The parameters of the plant are listed in Table \ref{param}. 
The Discrete-Time system matrices $(A_p,B_p,C_p,D_p)$ are obtained by discretizing the plant \eqref{power_AB}-\eqref{power_C} using zero-order hold with a sampling time $T_s=0.1\mbox{s}$. 

\begin{table}
\centering
\begin{tabular}{||c | c || c | c|| c | c ||} 
 \hline
 $K_{lm}$ & 1 & $T_{lm}$ & 6 &  $T_g$ & 0.2 \\
 \hline
 $T_{h}$ & $4$ & $T_s$ & 0.1 & $R$ & 0.05\\
 \hline
\end{tabular}
\caption{System Parameters}
\label{param}
\end{table}

The plant is stabilized locally with a static output feedback controller with constant gain $D_c=19$. The gains in \eqref{eq:cntrl} are obtained by minimizing a quadratic cost, using the MATLAB command \emph{dlqr}, resulting in:
\begin{align}
    K&=\begin{bmatrix}
        0.1986  & -0.0913  & -0.1143
    \end{bmatrix}\\
    L &= \begin{bmatrix}
       0.2735 &  -0.0509 & -0.2035
    \end{bmatrix}^\top.
\end{align}

\subsection{Initializing the mWM design algorithm}
We consider a mWM filter of state dimension $n_{\sigma}=2$. The mWM filter parameters are initialized as $A_q = 0.2I_2$, $B_q=0.7e_{2 \times 1}$, $C_q = 0.1 e_{1 \times 2}$, $B_h=0.2e_{2 \times 1}$, $C_h=0.05 e_{1 \times 2}$, $A_h=0.3I_2$, $D_q=0.15$, $D_h=0.1$ where $e_{a \times b}$ represents a unit matrix of size $a \times b$. 
The other mWM matrices are derived such that they satisfy \eqref{eq:WM:def}. 
All unspecified matrices are zero. Following Assumption~\ref{ass:param}, it is assumed that the filter parameters $\theta$ are known by the adversary. 
To ensure randomization, as mentioned in Theorem~\ref{thm:nonRepeat}, the parameters $D_h$ and $D_q$ are initialized in Algorithm~\ref{algo2} as random numbers within the range $[0.1,0.15]$. We fix the parameters of all the mWM filter parameters at their initial value except for the matrix $A_{\sigma}, \sigma \in \{q,w,h,g\}$, i.e., our aim is to find a diagonal $A_{\sigma}$ that minimizes the value of the AEC-OOG.

As discussed in Section~\ref{ch:design:algo}, $A_q$ and $A_h$ are matrices whose diagonal elements take values in $(-1,1)$.
The exhaustive search is performed with a grid size of $n_s = 0.3$, and the search is initialized with ${\epsilon_r = 1}, {\epsilon_a = 50}$. Furthermore, for numerical stability, we modify the objective function of \eqref{o1} to $\epsilon_r \gamma + \epsilon_a \gamma_a + \epsilon_p \mathrm{tr}(P)$, with ${\epsilon_p = 0.1}$.


\subsection{Result of Algorithm \ref{algo2}}
\begin{figure}
    \centering
    \includegraphics[width=7cm]{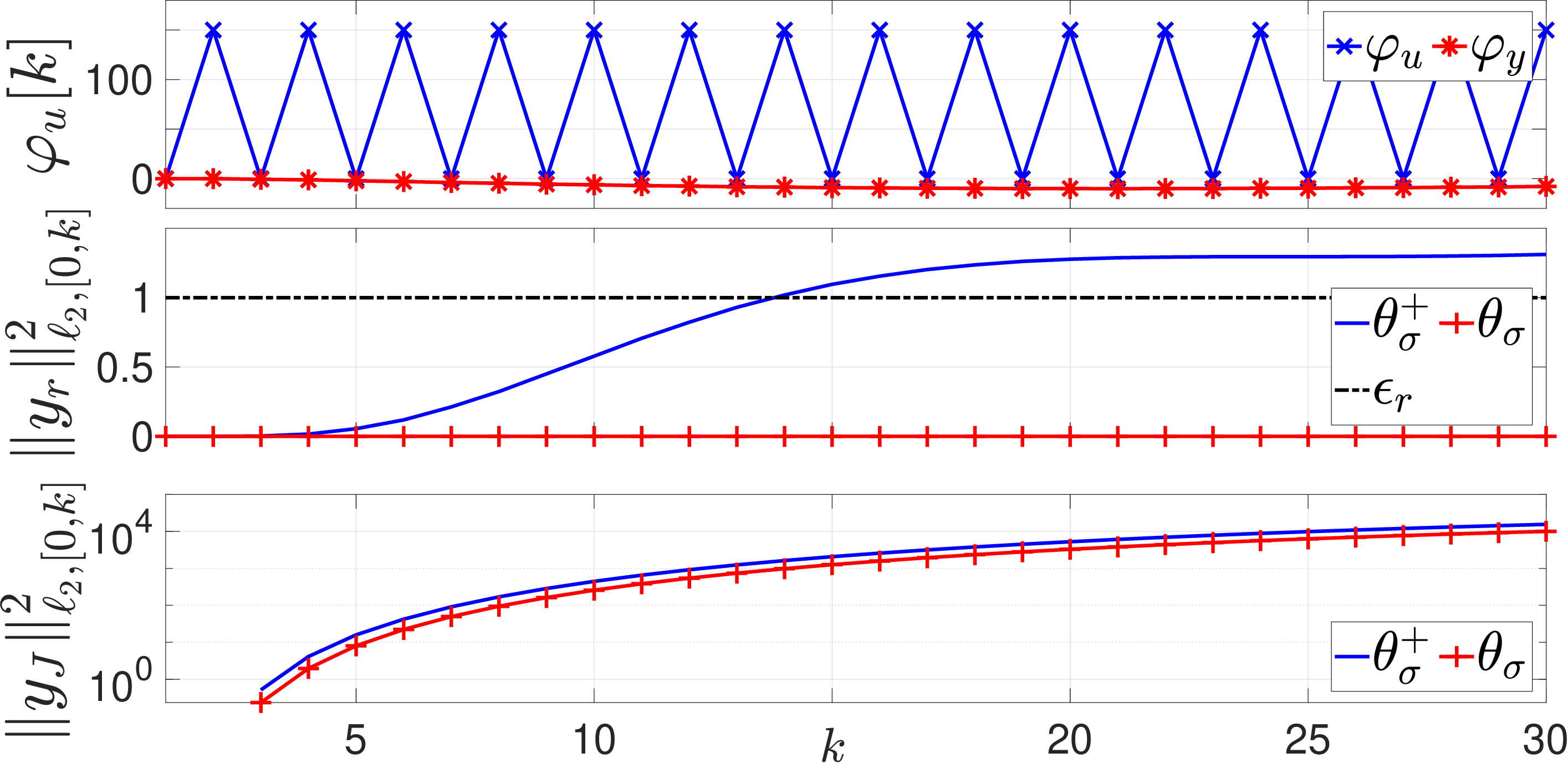}
    \caption{\ajg{(Top) The attack signal $\varphi_u$ in \eqref{eq:step} and its equivalent $\varphi_y$ from \eqref{eq:atk:cov}; (Middle) $\Vert y_r \Vert_{\ell_2,[0,k]}^2$, compared to $\epsilon_r$; (Bottom) $\Vert y_J \Vert_{\ell_2,[0,k]}^2$ before and after the mWM parameters are updated.}}
    \label{fig:no:switch}
\end{figure}

The optimal value of the matrices from the grid search are $A_q^* = -0.05I_2$ and $A_h^*=-0.65 I_2$. 
The corresponding value of $\mathcal{L}$ is $111.03$. The value of $D_q$ and $D_h$ were $0.1479$ and $0.1482$ respectively. The simulation is performed using Matlab 2021a with \textit{Yalmip} \citep{lofberg2004yalmip} and \textit{SDPT3v4.0} solver \citep{toh2012implementation}.
In the remainder, we compare the results obtained by repeated computation of Algorithm~\ref{algo2} compared to defining constant and random parameters.
Consider 
an adversary injecting the signals shown in Fig.~\ref{fig:no:switch},
\begin{equation}\label{eq:step}
    \varphi_u[k] = \begin{cases}
        150 & \text{if}\;k\;\text{mod}\;2=0 \\
        0 & \text{otherwise}
    \end{cases}
   \end{equation}
into the actuators, and $\varphi_y$ following \eqref{eq:atk:cov}.
    
\emph{Comparison with no parameter switching:}
The performance of the attack is shown in Fig.~\ref{fig:no:switch}, for the cases without switching and when switching happens at the attack onset with the optimal filter parameters.
Without switching $\theta$, although the performance is strongly degraded, the attack remains stealthy. Instead, if the mWM parameters are changed, it is detected after $15 \mbox{s}$.

\begin{figure}
    \centering
    \includegraphics[width=6.5cm]{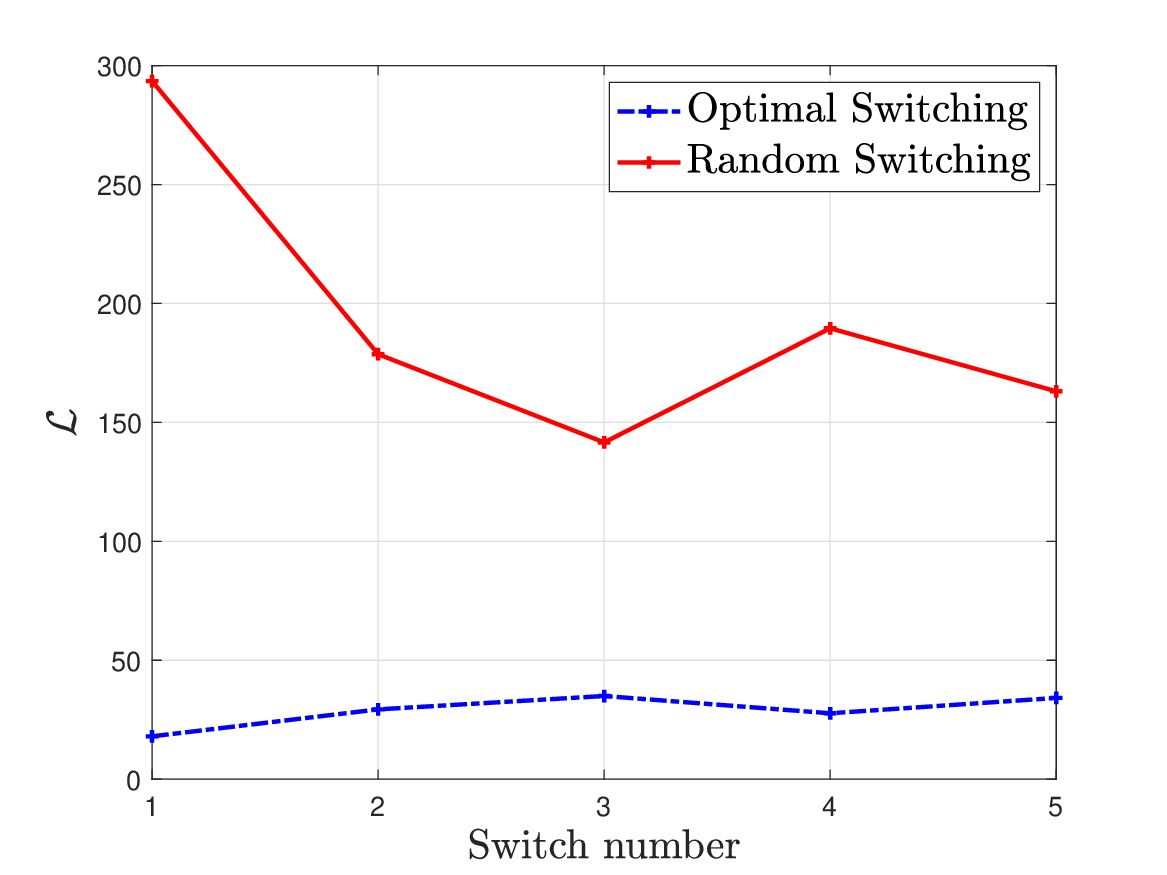}
    \caption{The values of $\mathcal{L}$ corresponding to the optimal and random values of the watermarking parameters.}
    \label{fig:switch}
\end{figure}

\emph{Comparison with random parameter switching:}
In this scenario, we suppose the mWM parameters are updated $5$ times, by running Algorithm~\ref{algo2}, and compared against $5$ random updates of $A_\sigma$ -- though their structure remains diagonal.
The results, shown in terms of values of $\mathcal L$ for both cases, are shown in Fig.~\ref{fig:switch}.
Here, the parameters of $D_h$ and $D_q$ are the same as used for selecting the optimal parameters. Since the parameters are not chosen optimally, the value of $\mathcal{L}$, the performance loss, is higher. 

\emph{Time complexity:}
To conclude, let us discuss thecomplexity of Algorithm~\ref{algo2}. All mWM parameters are fixed, apart from $A_\sigma$, which is a diagonal matrix of dimension $n_\sigma$, and, for each diagonal element of $A_\sigma$, $n_s$ points of the interval $(-1,1)$ are searched.
Given \eqref{eq:WM:def}, only $A_h$ and $A_q$ must be defined, while $A_w, A_g$ are defined algebraically; thus, define $n_\varsigma = n_q + n_h$.
The complexity of the algorithm grows both in $n_\varsigma$ and in $n_s$. Specifically: for $n_s = n_s^*$, the complexity is $\mathcal O(n_s^{*x})$; for $n_\varsigma = n_\varsigma^*$, the complexity is $\mathcal O(x^{n_\varsigma^*})$. Thus, the complexity is exponential in the choice of $n_\varsigma$ and polynomial in $n_s$.
We highlight that the average time of solution can be improved upon in two major ways. 
The first is via parallelization, as all SDPs can be solved independently; this provides a speed-up which depends on the number of compute nodes used to solve the problem.
The second method relies on reducing the number of SDPs to be solved, by removing those values of $A_h, A_q$ which do not lead to stable inverses, as defined by \eqref{eq:sys:inv}.

For the results presented here, a computer with an Intel Core i7-6500U CPU with 2 cores and 8GB RAM was used. The algorithm was run both with and without parallelization (parallelization was achieved by using Matlab's \texttt{parfor} command). Without parallelization, the algorithm took $384.25 \mathrm{s}$ to provide a result, whilst with parallelization this was $261.65 \mathrm{s}$, a  $31.9 \%$ speedup.

%% file: autosam.bbl
\begin{thebibliography}{31}
\providecommand{\natexlab}[1]{#1}
\providecommand{\url}[1]{\texttt{#1}}
\expandafter\ifx\csname urlstyle\endcsname\relax
  \providecommand{\doi}[1]{doi: #1}\else
  \providecommand{\doi}{doi: \begingroup \urlstyle{rm}\Url}\fi

\bibitem[Abdalmoaty et~al.(2023)Abdalmoaty, Anand, and Teixeira]{abdalmoaty2023privacy}
M.~R. Abdalmoaty, S.~C. Anand, and A.~M.~H. Teixeira.
\newblock Privacy and security in network controlled systems via dynamic masking.
\newblock \emph{IFAC-PapersOnLine}, 56\penalty0 (2):\penalty0 991--996, 2023.

\bibitem[Alisic et~al.(2023)Alisic, Kim, and Sandberg]{alisic2023model}
R.~Alisic, J.~Kim, and H.~Sandberg.
\newblock Model-free undetectable attacks on linear systems using lwe-based encryption.
\newblock \emph{IEEE Control Systems Letters}, 7:\penalty0 1249--1254, 2023.

\bibitem[Anand and Teixeira(2023)]{anand2023risk}
S.~C. Anand and A.~M.~H. Teixeira.
\newblock Risk-based security measure allocation against actuator attacks.
\newblock \emph{IEEE Open Journal of Control Systems}, 2:\penalty0 297--309, 2023.

\bibitem[Chen(1984)]{chen1984linear}
C.-T. Chen.
\newblock \emph{Linear system theory and design}.
\newblock Saunders college publishing, 1984.

\bibitem[Darup et~al.(2021)Darup, Alexandru, Quevedo, and Pappas]{darup2021encrypted}
M.~S. Darup, A.~B. Alexandru, D.~E. Quevedo, and G.~J. Pappas.
\newblock Encrypted control for networked systems: An illustrative introduction and current challenges.
\newblock \emph{IEEE Control Systems Magazine}, 41\penalty0 (3):\penalty0 58--78, 2021.

\bibitem[Dehnert et~al.(2021)Dehnert, Lerch, Grunert, Damaszek, and Tibken]{dehnert2021less}
R.~Dehnert, S.~Lerch, T.~Grunert, M.~Damaszek, and B.~Tibken.
\newblock A less conservative iterative {LMI} approach for output feedback controller synthesis for saturated discrete-time linear systems.
\newblock In \emph{2021 25th International Conference on System Theory, Control and Computing (ICSTCC)}, pages 93--100. IEEE, 2021.

\bibitem[Dinh et~al.(2011)Dinh, Gumussoy, Michiels, and Diehl]{dinh2011combining}
Q.~T. Dinh, S.~Gumussoy, W.~Michiels, and M.~Diehl.
\newblock Combining convex--concave decompositions and linearization approaches for solving {BMIs}, with application to static output feedback.
\newblock \emph{IEEE Transactions on Automatic Control}, 57\penalty0 (6):\penalty0 1377--1390, 2011.

\bibitem[Djouadi et~al.(2015)Djouadi, Melin, Ferragut, Laska, Dong, and Drira]{djouadi2015finite}
S.~M. Djouadi, A.~M. Melin, E.~M. Ferragut, J.~A. Laska, J.~Dong, and A.~Drira.
\newblock Finite energy and bounded actuator attacks on cyber-physical systems.
\newblock In \emph{2015 European Control Conference (ECC)}, pages 3659--3664. IEEE, 2015.

\bibitem[Escudero et~al.(2023)Escudero, Murguia, Massioni, and Zama{\"\i}]{escudero2023safety}
C.~Escudero, C.~Murguia, P.~Massioni, and E.~Zama{\"\i}.
\newblock Safety-preserving filters against stealthy sensor and actuator attacks.
\newblock In \emph{2023 62nd IEEE Conference on Decision and Control (CDC)}, pages 5097--5104. IEEE, 2023.

\bibitem[Fang et~al.(2019)Fang, Johansson, Skoglund, Sandberg, and Ishii]{fang2019two}
S.~Fang, K.~H. Johansson, M.~Skoglund, H.~Sandberg, and H.~Ishii.
\newblock Two-way coding in control systems under injection attacks: From attack detection to attack correction.
\newblock In \emph{Proceedings of the 10th ACM/IEEE International Conference on Cyber-Physical Systems}, pages 141--150, 2019.

\bibitem[Ferrari and Teixeira(2021)]{ferrari2020switching}
R.~M.~G. Ferrari and A.~M.~H. Teixeira.
\newblock A switching multiplicative watermarking scheme for detection of stealthy cyber-attacks.
\newblock \emph{IEEE Transactions on Automatic Control}, 66\penalty0 (6):\penalty0 2558--2573, 2021.
\newblock \doi{10.1109/TAC.2020.3013850}.

\bibitem[Gallo et~al.(2021)Gallo, Anand, Teixeira, and Ferrari]{gallo2021design}
A.~J. Gallo, S.~C. Anand, A.~M. Teixeira, and R.~M. Ferrari.
\newblock Design of multiplicative watermarking against covert attacks.
\newblock In \emph{2021 60th IEEE Conference on Decision and Control (CDC)}, pages 4176--4181. IEEE, 2021.

\bibitem[Griffioen et~al.(2020)Griffioen, Weerakkody, and Sinopoli]{griffioen2020moving}
P.~Griffioen, S.~Weerakkody, and B.~Sinopoli.
\newblock A moving target defense for securing cyber-physical systems.
\newblock \emph{IEEE Transactions on Automatic Control}, 66\penalty0 (5):\penalty0 2016--2031, 2020.

\bibitem[Hashemi and Ruths(2022)]{hashemi2022codesign}
N.~Hashemi and J.~Ruths.
\newblock Codesign for resilience and performance.
\newblock \emph{IEEE Transactions on Control of Network Systems}, 10\penalty0 (3):\penalty0 1387--1399, 2022.

\bibitem[Hemsley and E.~Fisher(2018)]{osti_1505628}
K.~E. Hemsley and D.~R. E.~Fisher.
\newblock History of industrial control system cyber incidents.
\newblock 12 2018.
\newblock \doi{10.2172/1505628}.
\newblock URL \url{https://www.osti.gov/biblio/1505628}.

\bibitem[Hu and Yan(2007)]{hu2007stability}
S.~Hu and W.-Y. Yan.
\newblock Stability robustness of networked control systems with respect to packet loss.
\newblock \emph{Automatica}, 43\penalty0 (7):\penalty0 1243--1248, 2007.

\bibitem[Lin et~al.(2023)Lin, Chong, and Murguia]{lin2023secondary}
Y.~Lin, M.~S. Chong, and C.~Murguia.
\newblock Secondary control for the safety of {LTI} systems under attacks.
\newblock \emph{IFAC-PapersOnLine}, 56\penalty0 (2):\penalty0 965--970, 2023.

\bibitem[Lofberg(2004)]{lofberg2004yalmip}
J.~Lofberg.
\newblock Yalmip: A toolbox for modeling and optimization in matlab.
\newblock In \emph{2004 IEEE {I}nternational {C}onference on {R}obotics and {A}utomation (IEEE Cat. No. 04CH37508)}, pages 284--289. IEEE, 2004.

\bibitem[Mo et~al.(2015)Mo, Weerakkody, and Sinopoli]{mo2015physical}
Y.~Mo, S.~Weerakkody, and B.~Sinopoli.
\newblock Physical authentication of control systems: Designing watermarked control inputs to detect counterfeit sensor outputs.
\newblock \emph{IEEE Control Systems Magazine}, 35\penalty0 (1):\penalty0 93--109, 2015.

\bibitem[Murguia et~al.(2020)Murguia, Shames, Ruths, and Ne{\v{s}}i{\'c}]{murguia2020security}
C.~Murguia, I.~Shames, J.~Ruths, and D.~Ne{\v{s}}i{\'c}.
\newblock Security metrics and synthesis of secure control systems.
\newblock \emph{Automatica}, 115:\penalty0 108757, 2020.

\bibitem[Park et~al.(2019)Park, Lee, Shim, Eun, and Johansson]{park2019stealthy}
G.~Park, C.~Lee, H.~Shim, Y.~Eun, and K.~H. Johansson.
\newblock Stealthy adversaries against uncertain cyber-physical systems: Threat of robust zero-dynamics attack.
\newblock \emph{IEEE Trans. on Automat. Contr.}, 64\penalty0 (12):\penalty0 4907--4919, 2019.

\bibitem[Sandberg et~al.(2022)Sandberg, Gupta, and Johansson]{sandberg2022secure}
H.~Sandberg, V.~Gupta, and K.~H. Johansson.
\newblock Secure networked control systems.
\newblock \emph{Annual Review of Control, Robotics, and Autonomous Systems}, 5:\penalty0 445--464, 2022.

\bibitem[Smith(2015)]{smith2015covert}
R.~S. Smith.
\newblock Covert misappropriation of networked control systems: Presenting a feedback structure.
\newblock \emph{IEEE Control Systems Magazine}, 35\penalty0 (1):\penalty0 82--92, 2015.

\bibitem[Stabile et~al.(2024)Stabile, Lucia, Youssef, and Franz{\`e}]{stabile2024verifiable}
F.~Stabile, W.~Lucia, A.~Youssef, and G.~Franz{\`e}.
\newblock A verifiable computing scheme for encrypted control systems.
\newblock \emph{IEEE Control Systems Letters}, 2024.

\bibitem[Teixeira et~al.(2015{\natexlab{a}})Teixeira, Sandberg, and Johansson]{teixeira2015strategic}
A.~Teixeira, H.~Sandberg, and K.~H. Johansson.
\newblock Strategic stealthy attacks: the output-to-output $\ell_2$-gain.
\newblock In \emph{2015 54th IEEE Conference on Decision and Control (CDC)}, pages 2582--2587. IEEE, 2015{\natexlab{a}}.

\bibitem[Teixeira et~al.(2015{\natexlab{b}})Teixeira, Shames, Sandberg, and Johansson]{teixeira2015secure}
A.~Teixeira, I.~Shames, H.~Sandberg, and K.~H. Johansson.
\newblock A secure control framework for resource-limited adversaries.
\newblock \emph{Automatica}, 51:\penalty0 135--148, 2015{\natexlab{b}}.

\bibitem[Toh et~al.(2012)Toh, Todd, and T{\"u}t{\"u}nc{\"u}]{toh2012implementation}
K.-C. Toh, M.~J. Todd, and R.~H. T{\"u}t{\"u}nc{\"u}.
\newblock On the implementation and usage of sdpt3--a matlab software package for semidefinite-quadratic-linear programming, version 4.0.
\newblock \emph{Handbook on semidefinite, conic and polynomial optimization}, pages 715--754, 2012.

\bibitem[Zhang et~al.(2023)Zhang, Gallo, and Ferrari]{zhang2023hybrid}
J.~Zhang, A.~J. Gallo, and R.~M. Ferrari.
\newblock Hybrid design of multiplicative watermarking for defense against malicious parameter identification.
\newblock In \emph{2023 62nd IEEE Conference on Decision and Control (CDC)}, pages 3858--3863. IEEE, 2023.

\bibitem[Zhang et~al.(2022)Zhang, Kasis, Polycarpou, and Parisini]{zhang2022sensor}
K.~Zhang, A.~Kasis, M.~M. Polycarpou, and T.~Parisini.
\newblock A sensor watermarking design for threat discrimination.
\newblock \emph{IFAC-PapersOnLine}, 55\penalty0 (6):\penalty0 433--438, 2022.

\bibitem[Zhou et~al.(1996)Zhou, Doyle, Glover, et~al.]{zhou1996robust}
K.~Zhou, J.~C. Doyle, K.~Glover, et~al.
\newblock \emph{Robust and optimal control}, volume~40.
\newblock Prentice hall New Jersey, 1996.

\bibitem[Zhu et~al.(2023)Zhu, Liu, Fang, Deng, and Cheng]{zhu2023detection}
H.~Zhu, M.~Liu, C.~Fang, R.~Deng, and P.~Cheng.
\newblock Detection-performance tradeoff for watermarking in industrial control systems.
\newblock \emph{IEEE Transactions on Information Forensics and Security}, 18:\penalty0 2780--2793, 2023.

\end{thebibliography}
